\documentclass{JHEP}

\newcommand{\be}{\begin{equation}}
\newcommand{\ee}{\end{equation}}

\newcommand{\bea}{\begin{eqnarray}}
\newcommand{\eea}{\end{eqnarray}}
\newcommand{\bean}{\begin{eqnarray*}}
\newcommand{\eean}{\end{eqnarray*}}

\def\beq{\begin{equation}}

\def\eeq{\end{equation}}

\preprint{ {\tt hep-th/yymmddd}}

\title{Seiberg Duality in Matrix Model}

\author{Bo Feng \\ Institute for Advanced Study \\
Einstein Drive, \\
Princeton, New Jersey, 08540 
}

\abstract{In this paper, we use the matrix model of pure
fundamental flavors (without the adjoint field) to check the 
Seiberg duality in the case of complete mass deformation.
We show that, by explicit integration at both sides of electric and
magnetic matrix models, the results agree with the prediction in
the field theory.}

\begin{document}

\paragraph{Motivation}

Since Dijkgraaf and Vafa gave the  matrix model conjecture
\cite{Dijkgraaf:2002fc,Dijkgraaf:2002vw,Dijkgraaf:2002dh},
there are a lot of related works to check, prove and 
generalize this conjecture, see \cite{Dorey:2002jc,Dorey:2002tj,
Dorey:2002pq,Chekhov:2002tf,Ferrari:2002jp,Fuji:2002wd,Dijkgraaf:2002pp,
Berenstein:2002sn,Gorsky:2002uk,Argurio:2002xv,McGreevy:2002yg,
Dijkgraaf:2002xd,Suzuki:2002gp,Ferrari:2002kq,Bena:2002kw,Demasure:2002sc,
Aganagic:2002wv,Gopakumar:2002wx,Naculich:2002hi,Cachazo:2002ry,
Dijkgraaf:2002wr,Tachikawa:2002wk}.
Especially in \cite{Dijkgraaf:2002xd} and \cite{Cachazo:2002ry},
 field theory arguments have been given to show why the calculation
of exact lower energy superpotential can be reduced to integration
in the corresponding matrix model.

The matrix model conjecture holds for more general cases than
the one proved in \cite{Dijkgraaf:2002xd,Cachazo:2002ry} although
the primary focus is still the theory with  adjoint fields. The 
generalization to  fundamental fields has been discussed
in \cite{McGreevy:2002yg,Argurio:2002xv,
Suzuki:2002gp,Bena:2002kw,Demasure:2002sc}. These kinds of generalizations
are desired because most applications in field theory 
will have  fundamental matters.
One particular interesting application is to see the Seiberg duality
\cite{Seiberg:1994pq} in the matrix model.

The standard example of Seiberg dual pair is following. On one
side, we have electric theory
$SU(N_c)$ with $N_f$ flavors $Q_j,\widetilde{Q}_j$ and no
superpotential. On another side, we have  magnetic theory
$SU(N_f-N_c)$ with $N_f$ 
flavors $q_j,\widetilde{q}_j$, meson field $X_i^j$ and
superpotential 
\be \label{dual-W}
W_{mag}={1\over \mu}X_i^j q_j  \widetilde{q}^i.
\ee
where $\mu$ is a dynamical scale \cite{Argyres} (equation (3.125)).
In general cases, these two theories will 
have flat directions in the moduli space, so when we try to
do the matrix model integration, we must take care of these zero
modes as did by Berenstein\cite{Berenstein:2002sn}. 
To avoid the complexity, we can deform
above theories by adding mass terms. For example, in the electric 
theory we add superpotential with non-degenerated
mass matrix\footnote{We can always redefine the 
field to bring the mass matrix into diagonal form.}
\be \label{electric}
W_{elec}= \sum_{j=1}^{N_f}  Q_j m_j \widetilde{Q}^j
\ee
Since we give all flavors nonzero mass
in the electric theory,
the IR field theory will be pure super Yang-Mills theory and we
know the exact effective action as
\be \label{YM}
W_{YM}= N_c (\hat{\Lambda}^{3N_c})^{1\over N_c}
\ee 
where $\hat{\Lambda}$ is the dynamical scale at IR and related to
dynamical scale $\Lambda$ at UV by
\be  \label{relations}
\hat{\Lambda}^{3N_c}=det(m) \Lambda^{3N_c-N_f}
\ee
The deformation in the electric theory will induce the
corresponding deformation in the magnetic theory as
\be \label{dualdefo}
W_{mag}={1\over \mu}X_i^j q_j  \widetilde{q}^i+ tr(Xm)
\ee

\paragraph{Matrix integration}

Now we will do the matrix integration for both electric and magnetic
theories. Let us do the electric theory first. 
The matrix model is 
\be
{1\over Vol(U(M))}\int \prod_{j=1}^{N_f} d Q_j dQ^{\dagger}_j
 e^{-{1\over g_s} \sum_{j=1}^{N_f} Q_j m_j
 Q^{\dagger j}}
\ee
where every $Q$ is a $M$ component vector and $\widetilde{Q}$ in 
the field theory has been treated as a conjugate of $Q$
\cite{Dijkgraaf:2002vw}. The part of flavor integration is
\bean
& &  \int \prod_{j=1}^{N_f} d Q_j dQ^{\dagger}_j
 e^{-{1\over g_s} \sum_{j=1}^{N_f} Q_j m_i
 Q_j^{\dagger}} 
 =   \prod_{j=1}^{N_f} ({\pi g_s \over m_j})^{M} \\
& = & (\pi g_s)^{M N_f} (det(m))^{-M} 
 =  e^{ M N_f \log(\pi g_s)- M\log(det(m))}
\eean
Except that, there is also the volume factor $e^{ {M^2\over 2} \log M}$.
Adding everything together we get
\bean
& & exp(  {M^2\over 2} \log M+ M N_f \log(\pi g_s)- M\log(det(m))\\
& = & exp( {1\over g_s^2} {S^2\over 2} \log {S\over g_s}+
{1\over g_s} [ S N_f \log(\pi g_s) - S\log(det(m))]) \\
& \equiv & exp( {1\over g_s^2} {\cal F}_{\chi=2} + 
{1\over g_s}{\cal F}_{\chi=1})
\eean
where we have grouped all term according to the genus expansion.
From the corresponds given in \cite{Dijkgraaf:2002dh,Argurio:2002xv}, 
we should identify
\be
-W_{elec}= N_c {\partial ({S^2\over 2} \log {S\over g_s})\over\partial S}
+ [ S N_f \log(\pi g_s) - S\log(det(m))]
\ee
In the matrix model, $S,M$ are dimensionless number, but in field
theory, $S$ is  dimension $3$. To match the field theory result,
we need to  replace $g_s$ by some proper dimensional number
heuristically. For example, we need to replace 
${S\over g_s}$ to ${S\over \Lambda^{3} e^{3/2}}$ to reproduce the
well known Veneziano-Yankielowicz term. Same heuristic argument tell us 
that $\pi g_s$ in the second term must be
replaced by dimension one number which can be chosen naturally as
 $\Lambda$\footnote{In fact, these results can be got by proper
definition of measure as in \cite{Ferrari:2002jp,Cachazo:2002ry}.}.
 Doing these replacements we get
\bean
-W_{elec} &= & 
N_c {\partial ({S^2\over 2} \log{S\over \Lambda^{3} e^{3/2}} )
\over\partial S}
+ [ S N_f \log(\Lambda) - S\log(det(m))] \\
& = & N_c[ S\log {S\over  \Lambda^{3}}-S]+
[ S N_f \log(\Lambda) - S\log(det(m))] \\
& = & N_c [S \log( {S \over (\Lambda^{3N_c-N_f} det(m))^{1\over N_c}})
-S]
\eean 
Minimized it we get 
$S= (\Lambda^{3N_c-N_f} det(m))^{1\over N_c}$ and
\be \label{electric_matrix}
W_{elec}= N_c (\hat{\Lambda}^{3N_c})^{1\over N_c}= 
N_c (\Lambda^{3N_c-N_f} det(m))^{1\over N_c}
\ee
which is  the famous Affleck-Dine-Seiberg superpotential for
pure $U(N)$ Yang-Mills gauge theory. Notice that the equation 
(\ref{relations}) is naturally shown in the matrix model.
Furthermore, in our calculation, we keep $N_f$ fixed while taking
$M\rightarrow \infty$.

Now we do the matrix model integration for the magnetic theory
\be
{1\over Vol(U(N))}\int dX \prod_j dq_j dq^{\dagger}_j exp(
{-1\over g_s}[  tr(m X)+ \sum_{i,j=1}^{N_f} {1\over \mu}
 X_i^j q_j  q^{\dagger i}])
\ee
where $X$ is $N_f\times N_f$ matrix and $q$ is $N$ component
vector. 
Doing the $X$ integration  first, we get a delta-function
\be
\delta(m_j^i+{q_j  q^{\dagger i} \over \mu})=\mu^{N_f^2} 
\delta(\mu m_j^i+q_j  q^{\dagger i})
\ee
Now the remainder part is exact the integration given in
\cite{Demasure:2002sc} ( equation (6) with $-\mu m$ taking the
place of $X$ there. Furthermore, since we always take large
$N$ limit with fixed $N_f$, there is not problem to apply
their result.), so we just cite  their result
\be \label{citeresult}
W_{mag}=(\widetilde{N}_c-N_f) ({\widetilde{\Lambda}^{3\widetilde{N}_c-N_f} 
\over det(-\mu m)})^{1\over \widetilde{N}_c-N_f}
\ee
where we use tilde to emphasize that it is in the magnetic theory.
Using the result $\widetilde{N}_c-N_f=-N_c$, 
$~~det(-\mu m)=(-)^{N_f} \mu^{N_f} det(m)$ and 
\be
\Lambda^{3N_c-N_f} \widetilde{\Lambda}^{3\widetilde{N}_c-N_f}
=(-)^{N_f-N_c} \mu^{N_f},
\ee
which can be found, for example,  in \cite{Argyres}.
we get immediately
\be \label{magnetic}
W_{mag}= N_c (\Lambda^{3N_c-N_f} det(m))^{1\over N_c}
\ee
where the $-$ sign in front of (\ref{citeresult}) is canceled 
exactly by factor $(-)^{N_c}$. Comparing (\ref{electric_matrix})
and (\ref{magnetic}) we see that they are same. Thus we give
an example to show how the matrix model can see the Seiberg duality.

We must emphasize that we did not realy derive the Seiberg duality from
the matrix model because the Seiberg duality holds in general situations
even without any mass deformation. We feel that to address the
full Seiberg duality, we must understand how to do the matrix integration
when there is flat directions in moduli spaces along the line 
\cite{Berenstein:2002sn}. Obviously, work should be generalized
to other generalized Seiberg duality, for example, the toric duality
addressed in  \cite{Feng:2000mi,Feng:2001xr,Feng:2001bn,Feng:2002zw,
Cachazo:2001sg,Beasley:2001zp,Berenstein:2002fi}. 

\paragraph{Acknowledgements}

We deeply thank  Nathan Seiberg for suggesting the mass deformation.
We also like to thank  Vijay Balasubramanian,
David Berenstein, Freddy Cachazo, Joshua Erlich, Yang-Hui He, 
Min-xin Huang,
Vishnu Jejjala and Asad Naqvi for fruitful discussion.
This research is supported under the NSF grant {\bf PHY-0070928}.

\bibliographystyle{JHEP}

\end{document}